\pdfoutput=1
\documentclass[journal]{IEEEtran}
\ifCLASSINFOpdf
\else
  \usepackage[dvips]{graphicx}
\fi

\usepackage{color,subfigure}
\usepackage{tikz}
\usetikzlibrary{intersections}
\usepackage{graphicx,amsmath,multirow}
\usepgflibrary{arrows} 
\usepgflibrary[arrows] 
\usetikzlibrary{arrows} 
\usetikzlibrary[arrows] 

\usepackage{fancyhdr}
\renewcommand{\footskip}{20 pt}

\begin{document}
%
\title{Self-field Effects and AC Losses in Pancake Coils Assembled from Coated Conductor Roebel Cables}
%
%
%

\author{Francesco~Grilli, Victor M. R. Zerme\~no, Enric~Pardo,~\IEEEmembership{Member,~IEEE}, Michal~Vojen\v ciak, J\"org~Brand, Anna~Kario, and Wilfried~Goldacker
\thanks{V. Zerme\~no, F. Grilli, M. Vojen\v ciak, J. Brand, A. Kario and W. Goldacker are with the Karlsruhe Institute of Technology. E. Pardo is with the Slovak Academy of Science, Bratislava, Slovakia.}
\thanks{Funding from the following sources is gratefully acknowledged: Helmholtz-University Young Investigator (Grant VH-NG- 617); EURATOM FU-CT-2007-00051 project co-funded by the Slovak Research and Development Agency under contract number DO7RP-0018-12.
}
\thanks{Manuscript received July 16, 2012}}

\maketitle

\begin{abstract}
In this contribution we develop a refined numerical model of pancake coils assembled from a coated conductor Roebel cable, which includes the angular dependence of the critical current density $J_c$ on the magnetic field and the actual (three-dimensional) shape of the current lead used to inject the current. Previous works of ours indicate that this latter has an important influence on the measured value of the AC losses. For the simulation of the superconductor, we used two alternative models based on different descriptions of the superconductor's properties and implemented in different mathematical schemes. For the simulation of the current lead we use a full three-dimensional finite-element model. The results of the simulation are compared with measurements and the main issues related to the modeling and the measurement of Roebel coils are discussed in detail.
\end{abstract}
\begin{IEEEkeywords}
Roebel cables, coils, AC losses, numerical simulations.
\end{IEEEkeywords}
\section{Introduction}
\IEEEPARstart{P}{ancake} 
coils are important for a wide variety of power applications employing HTS technology. Roebel cables made of rare-earth coated conductors constitute a promising example of a superconductor with high current capacity and low AC losses~\cite{Long:JOPCS10, Terzieva:SST10, Fleiter:SST13}. It is therefore natural to think of using Roebel cables for making pancake coils, and efforts in that direction have already begun~\cite{Grilli:TAS13a}. For optimizing the design of a coil, numerical models are very useful, as they can give important information on the distribution of magnetic field inside the coil as well as on the contributions to AC losses from different parts of the coil (for example, superconducting and metallic parts).

In a recent paper of ours we simulated differently sized pancake coils and computed their AC losses~\cite{Grilli:TAS13a}. The model utilized there had two important limitations: it did not take into account the dependence of the critical current density $J_c$ on the magnetic field  nor the actual geometry of the current leads used to inject the transport current. This latter has an important influence on the measured AC loss values.

In this paper, for the properties of the superconductor, we use an anisotropic angular dependence $J_c(B,\theta)$ derived from measurements. In addition, we simulate the real three-dimensional geometry of the current leads, which requires dedicated simulations.

\section{Coil geometry and Numerical Models}
\begin{figure}[t!]
\begin{center}
\includegraphics[width=0.9\columnwidth]{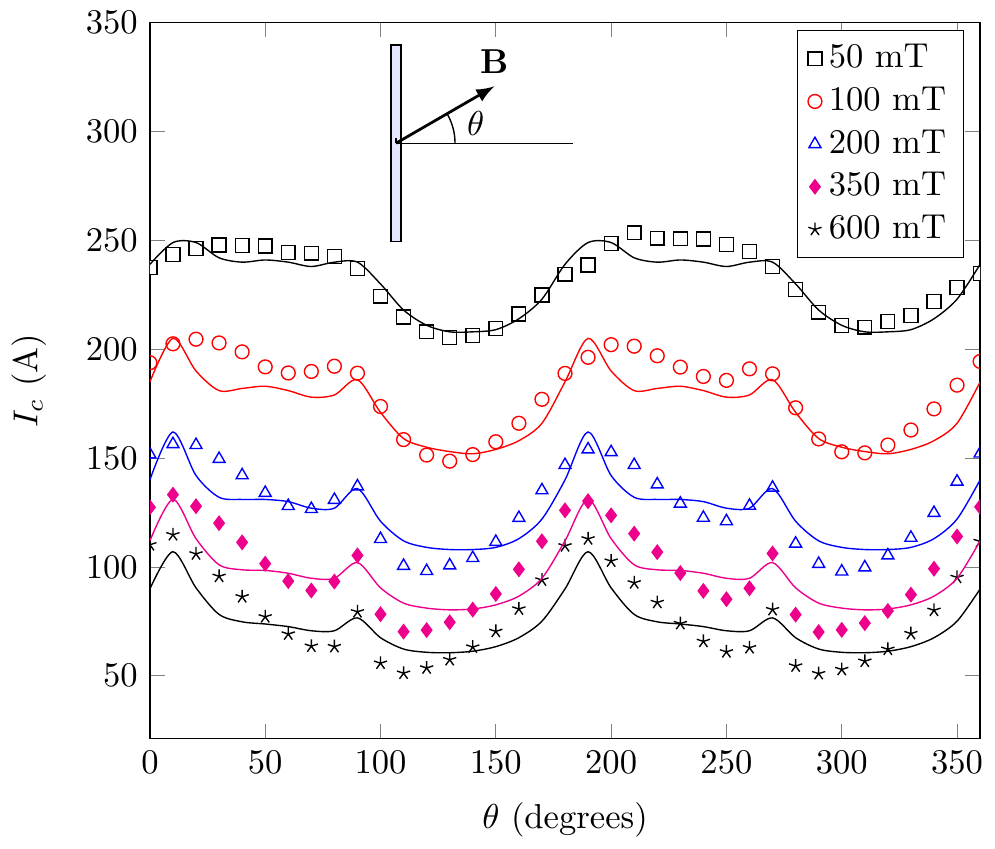}
\end{center}
\caption{Measured (symbols) and calculated (lines) angular dependence of $I_c$ on the applied magnetic field $H_a$. }\label{fig:IcBplot}
\end{figure}
\begin{figure}[t!]
\centering
\subfigure[]
{
\includegraphics[width=0.9\columnwidth]{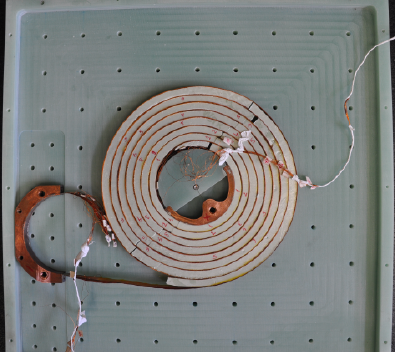}
\label{fig:Coil6}
}
\subfigure[]
{
\includegraphics[width=0.9\columnwidth]{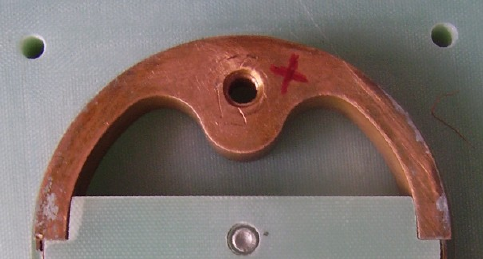}
\label{fig:Cu_contact}
}
\subfigure[]
{
\includegraphics[width=0.9\columnwidth]{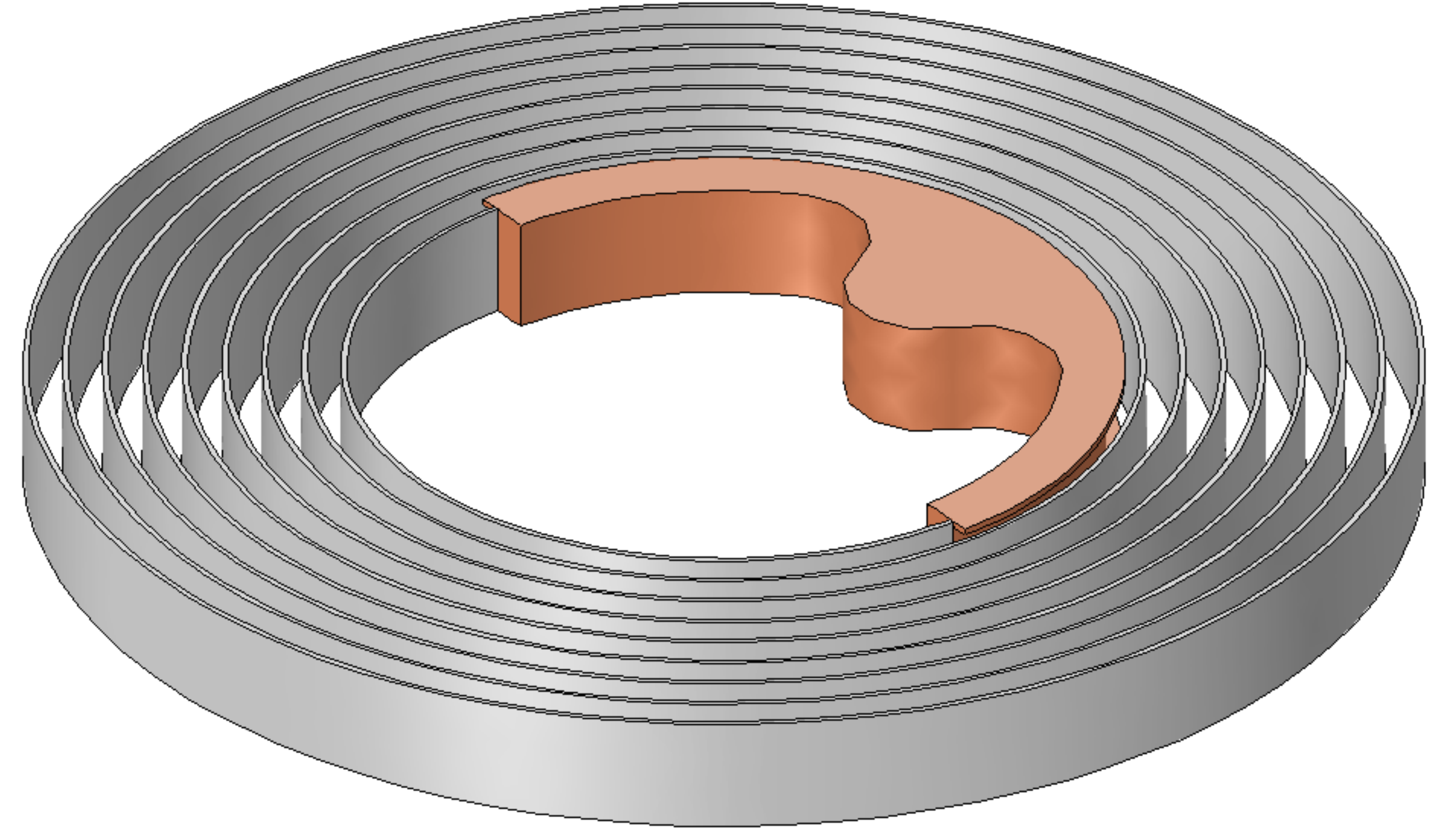}
\label{fig:CAD_4mm}
}
\caption{a) Pancake coil made of Roebel cable with experimental set-up for transport measurements; b) Detailed view of the inner copper contact used to inject the current; c) Simulated geometry for computing the AC losses in the current contact.}
\end{figure}

The considered coils are the same as those described in~\cite{Grilli:TAS13a}: they are assembled from one 5 meter long Roebel cable, composed of 10 strands obtained from 12 mm wide tape from Superpower, Inc. and with a DC self-field critical current of 936 A at 77 K. All the coils have an internal diameter of 10~cm. Their main properties are summarized in Table~\ref{tab:coils}. The coil with 10 mm spacing between the turns and the inner copper contact for injecting the current are shown in Fig.~\ref{fig:Coil6} and~\ref{fig:Cu_contact}, respectively. The shape of the copper contact was chosen with the objective of reducing eddy current losses (half-ring design) and of having a design suitable for the differently sized and shaped coils.
\begin{table}[ht!]
\renewcommand\arraystretch{1.3}
\caption{\label{tab:coils}Properties of the different coils.}
\centering
\begin{tabular}{llll}
Turns & Spacing (mm) &  ${ I_c}$ (A)&  \# of simulated strands \\
\hline \hline
13 & 0.1 & 456  & 130\\ \hline
9 & 4  & 661  & 90 \\ \hline
9 & 10  & 744  & 90\\ \hline
6 & 20  & 829  & 60 \\ \hline \hline
\end{tabular}
\end{table}

For the calculation of the AC losses in the superconductor, we used two axis-symmetric 2-D models considering only the coil's cross-section: the $H$-formulation of Maxwell equations with power-law resistivity for the superconductor and the Mimimum Magnetic Energy Variation model. Differently from~\cite{Grilli:TAS13a}, now a  $J_c(B,\theta)$ dependence is used and each Roebel strand is meshed with two elements along the thickness (instead of just one), so that the influence of the parallel component of the magnetic field can be accounted for. It is worth noting that, due to the asymmetric anisotropy of the utilized $J_c(B,\theta)$, it is not possible to consider only one half of the coil's cross-section, as it was done in the case of the constant $J_c$. Therefore, the numbers in the last column of Table~\ref{tab:coils} represent the actual number of simulated tapes. 

The $J_c(B,\theta)$ dependence was derived from the experimental data $I_c(H_a,\theta)$ on individual tapes, where $I_c$ is the tape's critical current (determined with the 1 $\mu$V/cm criterion), $H_a$ is the amplitude of the applied field, and $\theta$ its orientation with respect to the tape -- see Fig.~\ref{fig:IcBplot}. Due to the self-field effects occurring at low-fields, $J_c(B,\theta)$ cannot be extracted by a simple fit of the $I_c(H_a,\theta)$ data, but by means of dedicated simulations, as explained in~\cite{Vojenciak:SST11}.

The utilized $J_c(B,\theta)$ has the following analytical form:
\begin{equation}\label{eq:JcBtheta}
J_c(B, \theta) =(J_{c_{90}}^m+J_{c_{10}}^m+J_ {c_{50}}^m)^{1/m},
\end{equation}
which represents a superposition of three elliptical dependences peaked at different angles. (See~\cite{Pardo:SST11} for details.) The power exponent $m=8$ has the function of determining which of the three is dominant at a given angle of the local magnetic flux density. In Eq.~\eqref{eq:JcBtheta} each $J_{c_x}$ component  ($x$=90, 10, 50) is given by
\begin{equation}
J_{c_{x}}=	J_{c0_{x}}/[1+(B  f_{x}/B_{0_{x}})^{\beta_{x}}]
\end{equation}
where $B$ is the magnitude of the local flux density and the coefficients $\beta_{x}$, $d_x$ and $u_x$ are given in Table~\ref{tab:coefficients}.
\begin{table}[ht!]
\renewcommand\arraystretch{1.3}
\caption{\label{tab:coefficients}Coefficients of the $J_c(B, \theta)$ dependence.}
\centering
\begin{tabular}{lllll}
$x$ 	& $J_{c0_{x}}~ {\rm (A/m^2)}$	& $\beta_{x}$	&  $d_x$ (rad)	& $u_x$ \\ \hline \hline
90	& 3.2 $\cdot 10^{10} $  & 0.035		& 1.5708  & 6  \\ \hline 
10	& 3.5 $\cdot 10^{10} $ &	0.02		& 0.1745  & 7  \\ \hline 
50	& 3.5 $\cdot 10^{10} $ &	0.032	& 0.8727  & 1.5  \\ \hline \hline 
\end{tabular}
\end{table}

The simulation of the current contact requires a 3-D model,
and it is therefore carried out separately.
In this case, we are not interested in the details of the
superconducting strands composing the coil, but rather on the
effect of the coil on the copper contact. Therefore, each coil is simulated as a set of concentric rings, where each ring carries the same current. This provides a realistic model that takes into account that the current lead is exposed to both transport current and applied magnetic field, the latter provided from the coil itself. In this way, losses in the copper lead due to eddy currents are estimated. The geometric model of the lead was taken from the same computer-aided design file used to build the lead itself, with the exception that the hole in the center was omitted as it would be filled once the contact is inserted -- see Fig.~\ref{fig:CAD_4mm}.

The total AC losses of the coil are computed by adding the losses of the superconductor (from the axis-symmetric 2-D model) to the losses in the current lead (3-D model), which are essentially caused by the eddy currents induced by the magnetic field generated by the coil. In both cases, the losses are computed by integrating the  power density ${\mathbf J} \cdot {\bf E}$ over the volume of interest and averaging over the second half-cycle:
\begin{equation}
Q=\frac{2}{T}\int\limits_{T/2}^{T} \int\limits_{\Omega}{\mathbf J} \cdot {\bf E} ~{\rm d}\Omega {\rm d}t,
\end{equation}
where $T$ is the period and $\Omega$ the superconductor or copper domain.

\section{Results}
\begin{figure}[h!]
\centering
\subfigure[]
{
\includegraphics[width=0.95\columnwidth]{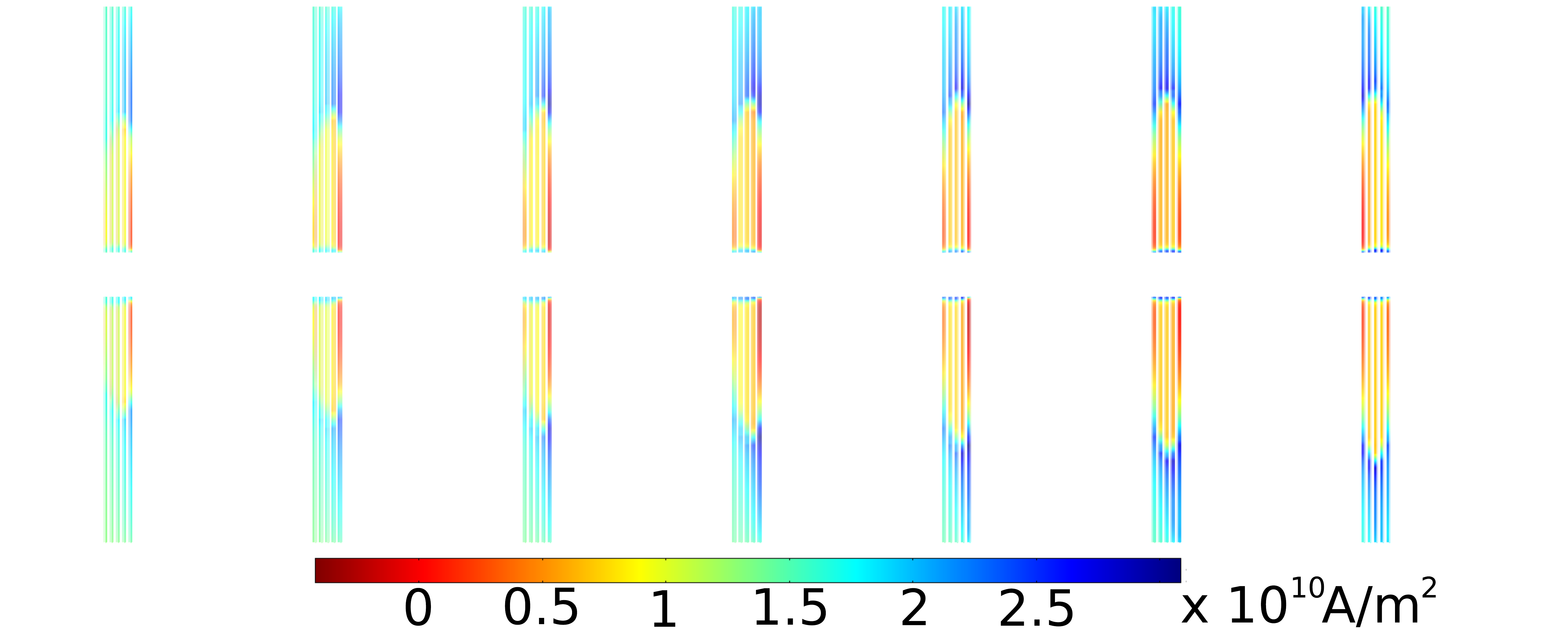}
\label{fig:J_final}
}
\subfigure[]
{
\includegraphics[width=0.95\columnwidth]{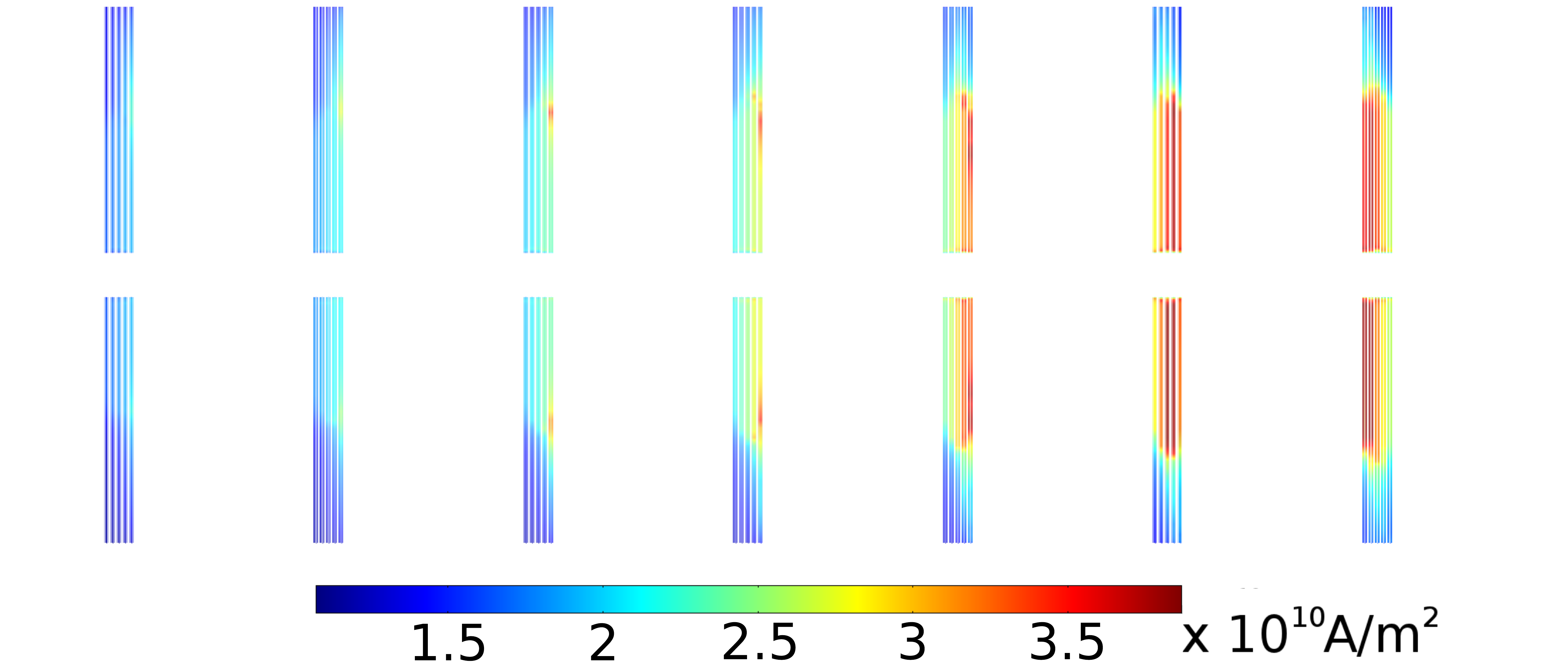}
\label{fig:Jc(B)_final}
}
\subfigure[]
{
\includegraphics[width=0.9\columnwidth]{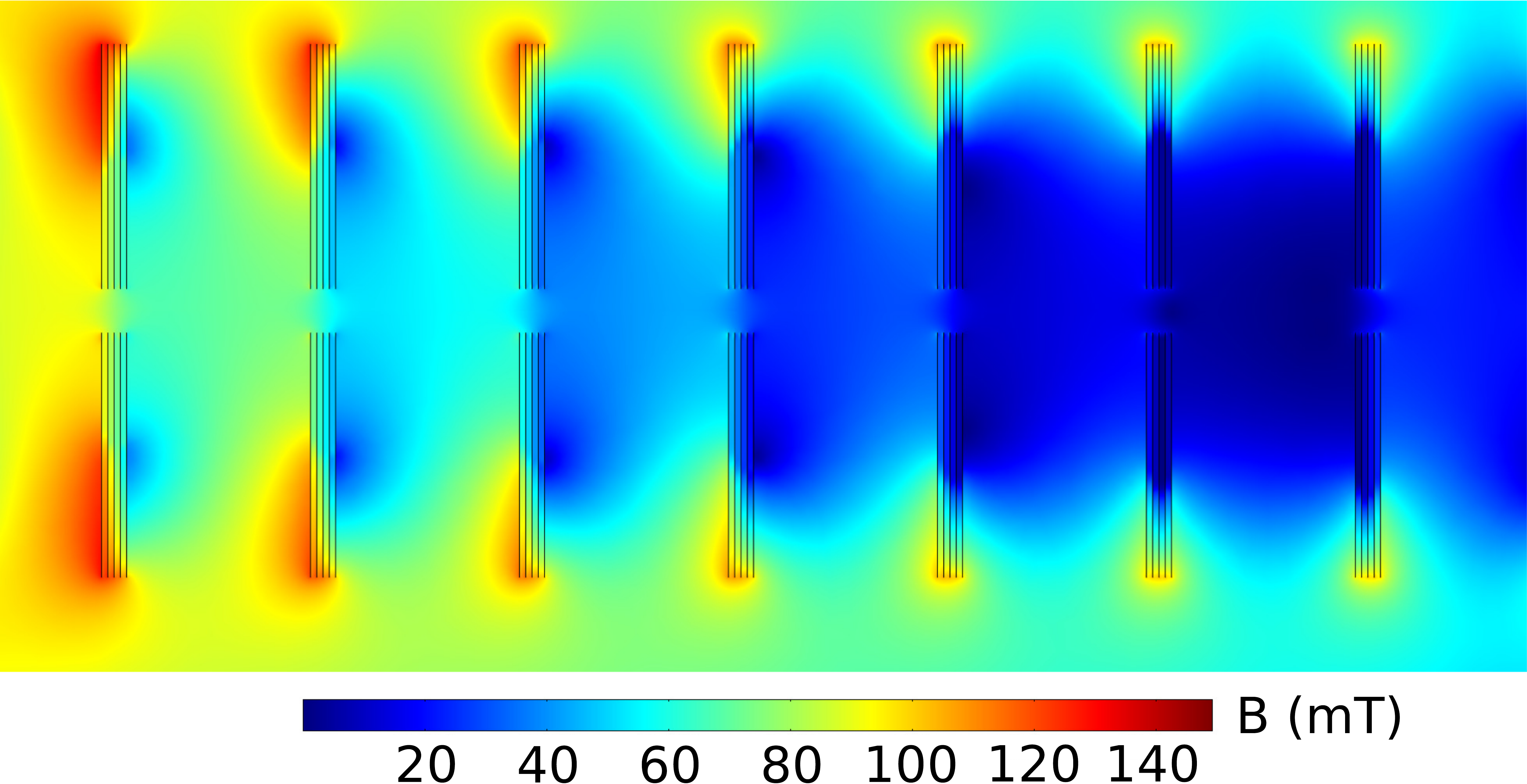}
\label{fig:normB(T)_final}
}
\subfigure[]
{
\includegraphics[width=0.95\columnwidth]{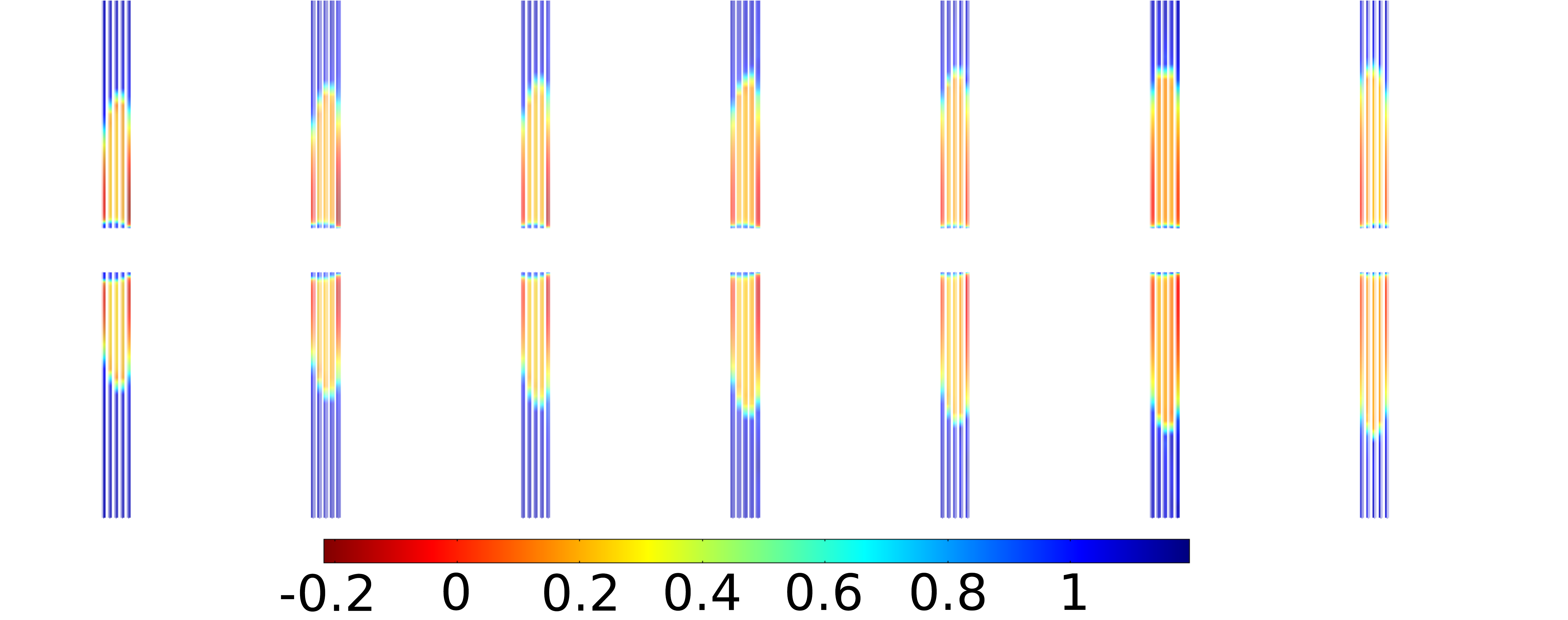}
\label{fig:Jnorm_final}
}
\subfigure[]
{
\includegraphics[width=0.95\columnwidth]{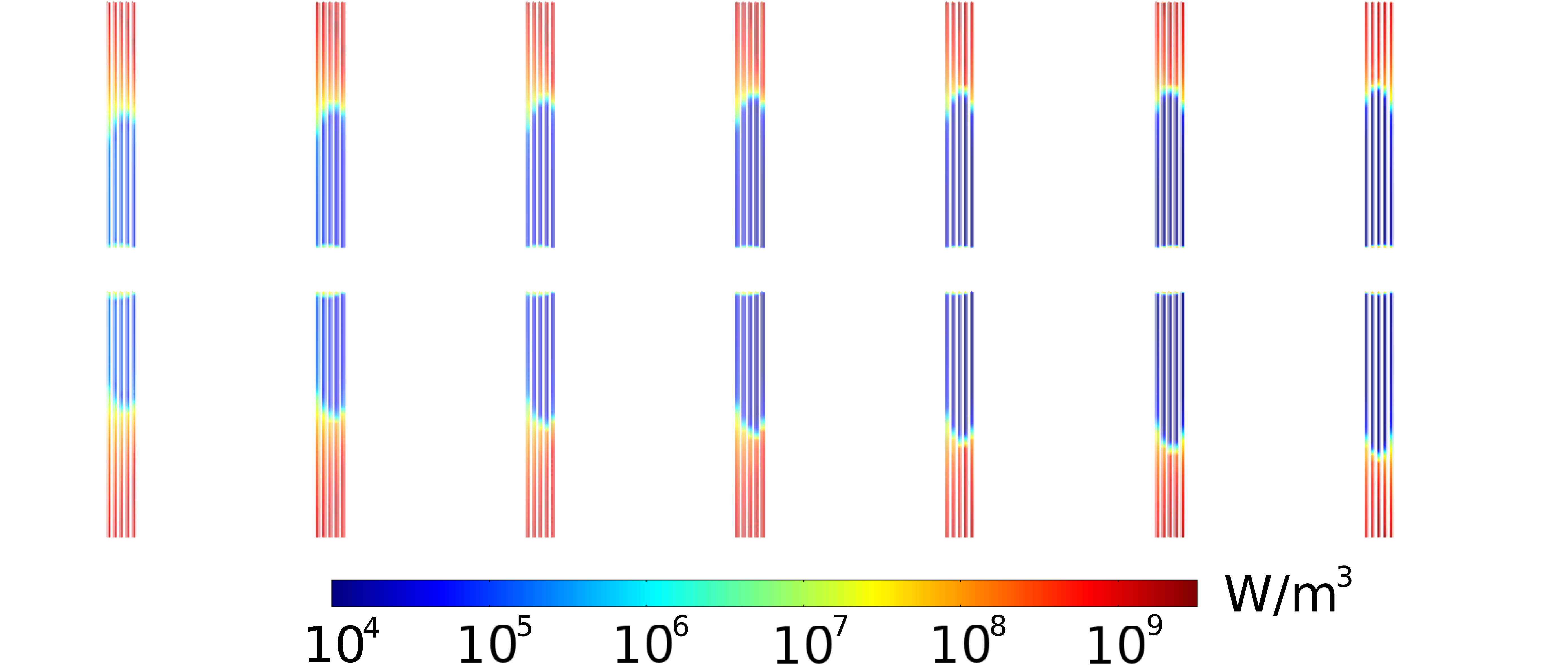}
\label{fig:log10(time_average(EJ))_final}
}
\caption{Current, field, and loss distributions inside the coil with 4 mm turn spacing at the peak of the current: a) Current density $J$; b) Critical current density $J_c(B,\theta)$; c) Magnetic flux density; d) Normalized current density $J/J_c(B,\theta)$; e)  Average power density per cycle. Frequency of the current: 50 Hz.}
\end{figure}
The fact that $J_c$ depends on the magnetic field influences the current density distribution: the maximum of the current does not occur at the edges of the coils, but slightly at their interior. Since the dissipation occurs where $J$ is the highest and exceeds the local value of $J_c$, this means that also the distribution of the local dissipation has a similar pattern. This is clearly visible in Figs.~\ref{fig:J_final} and \ref{fig:Jc(B)_final}, which represent the distributions of the current density and of the reduction of the critical current density (according to Eq.\eqref{eq:JcBtheta}), respectively. In those figures, which refer to the coil with 4 mm turn separation, the center of the coil is on the left side. The transport current is 661 A and the distributions are taken at the current peak. It can be seen that the maximum of the current density does not occur at the edges of the coil, but at its interior. This is because at the edges $J_c$ decrease quite strongly due to the higher  magnetic field there.

The magnetic field is higher close to the coil axis (left side of figures) than in its outer region (right side of figures) -- see Figs.~\ref{fig:normB(T)_final}. As a consequence, the internal turns have much lower current carrying capability, as in the case of pancake coils made of a single coated conductor~\cite{Gomory:TAS13}. The local critical current density is about $2 \cdot \rm 10^{10} A/m^2$ (or less) in the innermost turn, but it reaches and exceeds $3 \cdot \rm 10^{10} A/m^2$ in most parts of the outermost turns.

Figure~\ref{fig:Jnorm_final} shows the distribution of $J/J_c(B,\theta)$, i.e. of the locally normalized current density. This quantity is very informative because it tells us where most of the dissipation occurs: in the regions were the {\it local} value of $J_c$ is reached (and possibly exceeded). The other regions, characterized by current densities lower than the local $J_c$ value, do not practically contribute to the AC losses. This is visualized in Fig.~\ref{fig:log10(time_average(EJ))_final}, which displays the local dissipation averaged over the second half-cycle.
\begin{figure}[t!]
\begin{center}
\includegraphics[width=0.9\columnwidth]{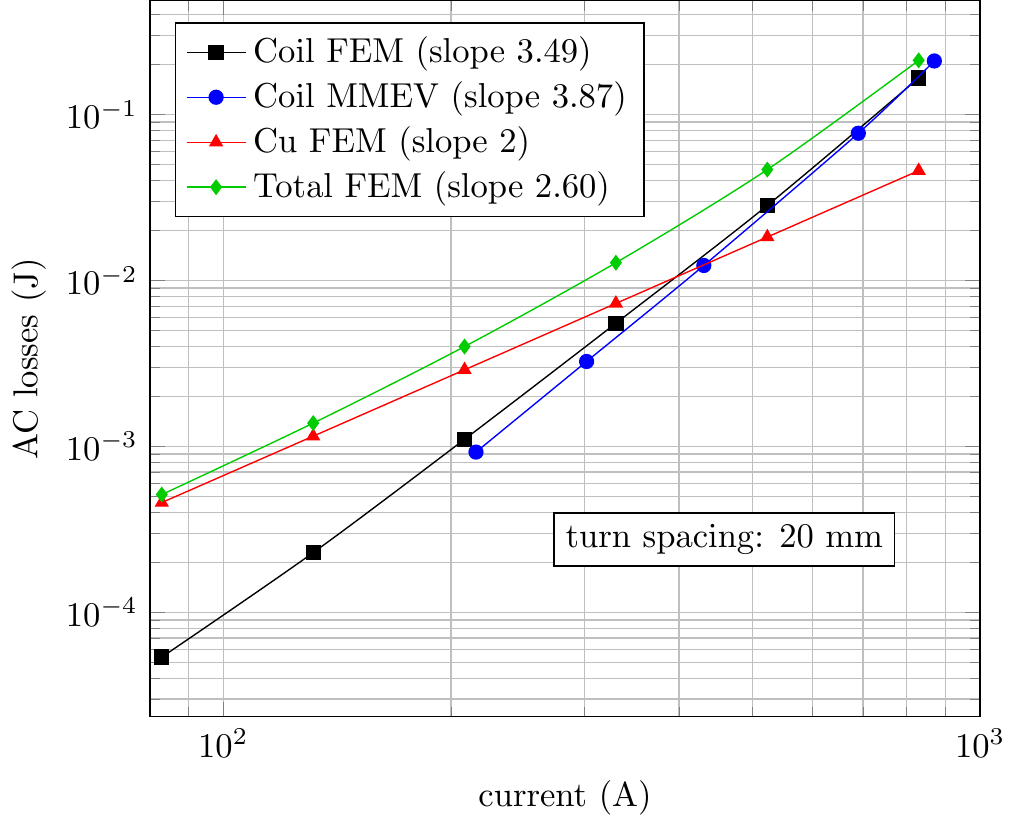}
\end{center}
\caption{Comparison of the AC losses of the 20 mm coil computed with FEM and MMEV. The losses in the contact and the total losses are also shown.}
\label{fig:slopes}
\end{figure}
\begin{figure}[ht!]
\begin{center}
\includegraphics[width=0.9\columnwidth]{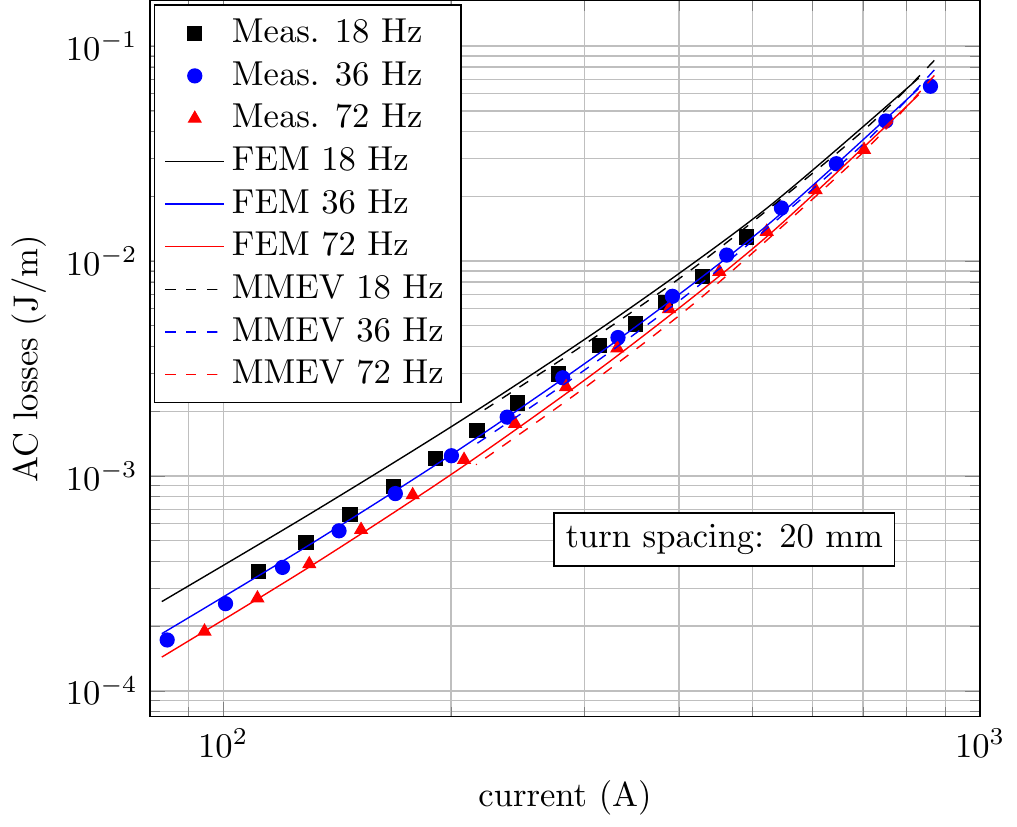}
\end{center}
\caption{Measured and calculated transport AC losses in the coil with 20 mm separation between the turns.}
\label{fig:Qcoil_20mm_plot}
\end{figure}
\begin{figure}[ht!]
\begin{center}
\includegraphics[width=0.9\columnwidth]{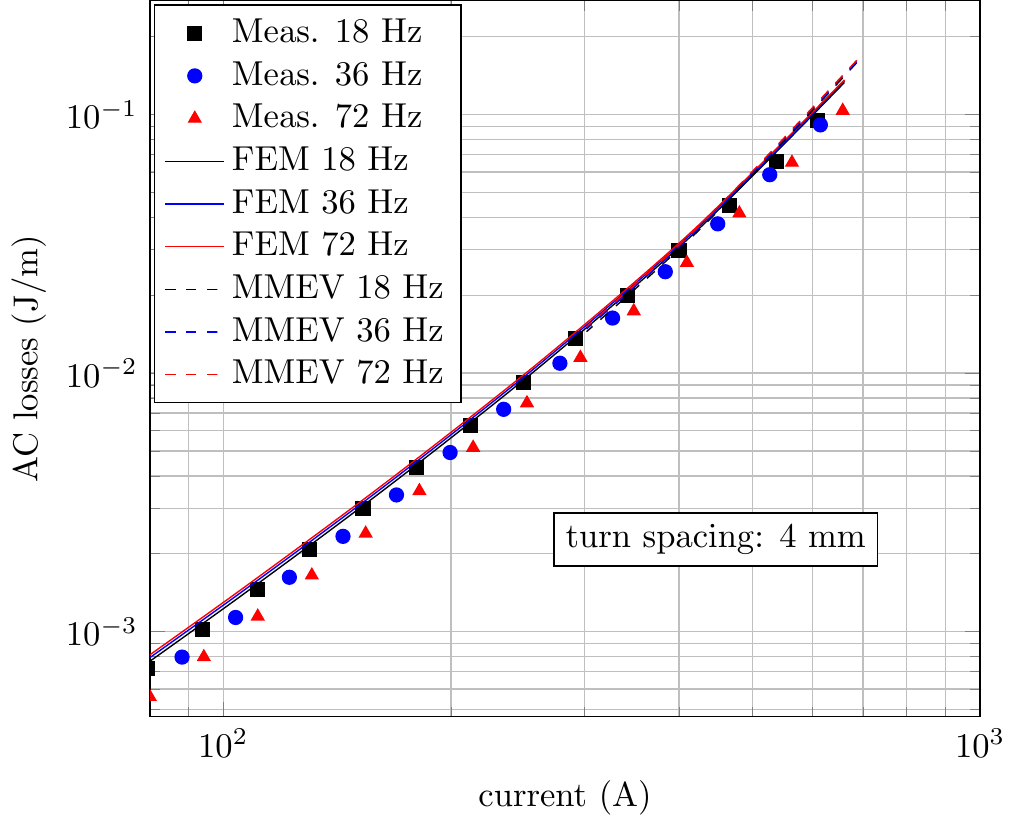}
\end{center}
\caption{Measured and calculated transport AC losses in the coil with 4 mm separation between the turns.}
\label{fig:Qcoil_4mm_plot}
\end{figure}

The AC losses as a function of current at different frequencies were calculated and compared to measurements. 
Figure~\ref{fig:slopes} shows a comparison of the AC losses of the 20 mm coil computed with FEM and MMEV, as well as the corresponding losses in the copper contact. The agreement between the model is good, especially at medium and high currents. The slight discrepancy at low current is most probably caused by the relatively coarse utilized mesh (only two elements along the thickness) and by the very little current penetration inside the superconductor.  The losses in the superconductor increase with a power of the transport current higher than 3, more precisely 3.49 and 3.87 for the FEM and MMEV models, respectively. This is different from the measured data, which increase with smaller power, typically around 2.5. This is due to the losses in the copper contact, which, being of eddy current nature, increase with the power 2 of the transport current. At low currents, these losses are larger than the superconductor losses, which on the contrary become dominant at high current. The crossing of the two loss curves explains why the total losses increase with an intermediate power of the current.

Two exemplary AC loss plots for the coils with 20 and 4 mm separation between the turns are shown in Figs.~\ref{fig:Qcoil_20mm_plot} and~\ref{fig:Qcoil_4mm_plot}, respectively. The agreement between simulations and measurements is very good, which confirms the reliability of our models for computing AC losses in devices made of coated conductor Roebel cables. These results also indicate that for the calculation of AC losses in the superconductor at normal operating currents a 2-D model is sufficient, and complex 3-D models of Roebel cables as the one presented in~\cite{Zermeno:SST13} are only necessary to study very local effects, like the dissipation in critical zone during overcritical excursions. On the other hand, the model of the current lead, while being fully 3-D, does not add an excessive complexity due to the relatively simple geometry considered and the simulation of a linear material.

In our case of a pancake coil in self-field, the 3-D simulations do not differ from the 2-D ones because the loss in the crossing parts is practically zero. The cause is that the magnetic field in the crossing parts is parallel to the superconductor surface. In a stack of pancakes or a solenoid there is a significant perpendicular component of the magnetic field in the cable center. Then, the crossing parts will present magnetization loss. 3-D simulations under uniform applied field show that the contribution of the crossing part is important at low applied fields. Therefore, for stacks of pancakes or solenoids, 3-D and 2-D simulations will differ at low current amplitudes.
\section{Conclusion}
With this work we modeled pancake coils made of Roebel cables using an  2-D axis-symmetric model for the superconductor (which includes the anisotropic angular dependence of the critical current density) and a three-dimensional model of the copper current leads used in experiments. The simulation of the superconductor was carried out using two different approaches -- the $H$-formulation of Maxwell equations with power-law resistivity for the superconductor and the Minimum Magnetic Energy Variation model. These models gave results in very good agreement with each other, and proved to be two complementary methods for computing losses in superconductors applications.

The angular anisotropy of the superconductor's $J_c$ strongly influences the way current penetrates inside the coil: the most remarkable features are that the maximum of the current density occurs somewhere at the interior of the strands (and not at their edges) and that the current density distribution across the coil's cross-section does not present symmetries.

Most of the dissipation occurs in the parts of the strands facing the exterior of the coil, the internal part being screened from field penetration and essentially loss-free. The extent of field penetration is strongly influenced by the shape of the coil and by the number of turns: the tighter the turns are wound, the higher the losses are.

The experimentally measured losses are influenced by the copper contact, more precisely by its eddy current losses caused by the self-field of the coil. The influence of the copper contact is particularly important for low transport current, where the losses in the copper are higher than those in the superconductor. Our numerical simulations are able to reproduce the experimental observations very well and they constitute a very useful tool for optimizing the design of coils made of coated conductor Roebel cables, and more generally, of any kind of superconductors.

\section*{Acknowledgment}
V. Zerme\~no would like to thank Dr. Mads P. S\o rensen (Department of Applied Mathematics at the Technical University of Denmark) for the access to the computational resources.


\end{document}